# Electrically Tunable Magnetoconductance of Close-Packed CVD Bilayer Graphene Layer Stacking Walls

Qicheng Zhang[1, †], Sheng Wang[2, †], Zhaoli Gao[3, †], Sebastian Hurtado-Parra[1], Joel Berry[4], Zachariah Addison[1,5], Paul Masih Das[1,6], William M. Parkin[1], Marija Drndic[1], James M. Kikkawa[1], Feng Wang[7,8,9], Eugene J. Mele[1,*], A. T. Charlie Johnson[1,*], Zhengtang Luo[10,*]

[1] Department of Physics and Astronomy, University of Pennsylvania, Philadelphia 19104, USA

[2] School of Physics and Technology, Wuhan University, Wuhan 430072, P. R. China

[3] Department of Biomedical Engineering, The Chinese University of Hong Kong, Shatin, Hong Kong

[4] Materials Science Division, Lawrence Livermore National Laboratory, Livermore, CA 94550, USA.

[5] Department of Physics, The Ohio State University, Columbus, OH 43210, USA.

[6] Department of Materials Science & Engineering, Northwestern University, Evanston, IL 60208, USA

[7] Department of Physics, University of California at Berkeley, Berkeley, CA, USA

[8] Materials Science Division, Lawrence Berkeley National Laboratory, Berkeley, California 94720, USA.

[9] Kavli Energy NanoSciences Institute at the University of California, Berkeley and the Lawrence Berkeley National Laboratory, Berkeley, California, 94720, USA.

[10] Department of Chemical and Biological Engineering, Hong Kong University of Science and Technology, Clear Water Bay, Kowloon, Hong Kong

[†]These authors contributed equally to this work.

## Abstract

Quantum valley Hall (QVH) domain wall states are a new class of one-dimensional (1D) one-way conductors that are topologically protected in the absence of valley mixing. Development beyond a single QVH channel raises important new questions as to how QVH channels in close spatial proximity interact with each other, and how that interaction may be controlled. Scalable epitaxial bilayer graphene synthesis produces layer stacking wall (LSW) bundles, where QVH channels are bound, providing an excellent platform to study QVH channel interactions. Here we show that distinct strain sources lead to the formation of both well-separated LSWs and close packed LSW bundles. Comparative studies of electronic transport in these two regimes reveal that close-packed LSW bundles support electrically tunable magnetoconductance. The coexistence of different strain sources offers a potential pathway to realize scalable quantum transport platform based on LSWs where electrically tunability enables programmable functionality.

## Introduction

Two-dimensional materials with a honeycomb lattice structure, such as graphene or transition metal dichalcogenides in the 1H phase, have two inequivalent valleys in k-space whose states are degenerate in energy and can be indexed by a valley degree of freedom. The inversion symmetry of bilayer graphene can be broken by application of an electric field along the layer normal, leading to an energy gap and concentrated momentum space Berry curvatures in each valley with opposite signs. The momentum space integral of the Berry curvature about each valley can be associated to a topological charge[1,2]. QVH domain wall states are confined modes bound to domain walls where valley-dependent topological charge changes its sign, and these chiral boundary modes are topologically protected in the absence of valley mixing, leading to long mean free paths for these states[2–7]. Inter-wall coupling of QVH domain wall states occurs when wavefunctions on neighboring walls overlap[8–11]. Although phenomena such as current partitioning[8,9] and valley filtering[10] have been observed[11] in a point contact geometry of electric field walls, the control of the inter-wall coupling has not yet been demonstrated. The region between two inequivalent Bernal-stacking orders contains a layer stacking wall (LSW) where QVH domain wall states are bound[6,12,13]. Rich LSW patterns have been reported in scalable synthesized bilayer graphene through epitaxial growth[14,15], including bundles, a natural assembly of 1D systems[16,17]. The combination of scalability and much simpler electronic device design than electric field walls makes LSWs in epitaxial bilayer graphene an excellent platform to study and apply QVH domain wall states and their coupling. However, LSWs are strain solitons that interact elastically. Adjacent pairs can merge and annihilate or stay intact depending on their separation and the relative character of their elastic fields[14]. This complex strain problem creates challenges to develop the LSW-based QVH platform. Here, we report markedly different magnetoconductance (MC) behaviors for well-separated LSWs compared to their close-packed counterparts. Substrate corrugation during growth induces proximity-avoiding, well-separated LSWs, whose positive MC response is small and insensitive to the variation of perpendicular electrical field. In contrast, localized stresses promote the formation of close-packed LSW bundles, where the interaction of Bloch electrons with an external magnetic field leads to a zero-magnetic field MC minimum whose depth can be tuned with an external perpendicular electric field. These phenomena can be explained by a model of electrically tunable domain wall mode wavefunction overlap for the case of close-packed LSWs.

## Results and Discussion

Bilayer graphene flakes used in these experiments were grown by chemical vapor deposition (CVD) on an optimized Ni-Cu gradient alloy substrate[18]. Analogous to recent discoveries concerning ABC/ABA domain walls in trilayer graphene[19], and in contrast to what has been observed earlier in epitaxially grown bilayer graphene flakes[14,15], the predominant LSWs are induced by substrate corrugation (Supplementary Note 1), and the AB/BA stacking domains appear as stripes that map well (method see Supplementary Note 2) onto substrate terraces (Fig. 1a). In dark field transmission electron microscopy (DF-TEM), zig-zag LSWs, which extend along the graphene zig-zag direction, are clearly visible in the image of the $(2\bar{1}\bar{1})$ reflection (Fig. 1b). In contrast, the LSWs are no longer apparent when the $(1\bar{2}1)$ reflection is used, in agreement with the invisibility criterion[15], since the direction of the Burgers vector (yellow arrows at the lower part of Figs. 1b-c) is perpendicular to the diffraction vector for this case. This observation indicates the existence of strain coherence across these LSWs. Armchair LSWs also show such coherence (Supplementary Figs. 3a-d). In contrast to LSWs created by mechanical forces on a flat substrate, where strain is dominated by shear force[20], the coherent strain of these LSWs is dominated by tensile/compressive forces, following the maximum curvature direction of the peaks and valleys on the substrate (lower part of Fig. 1a).

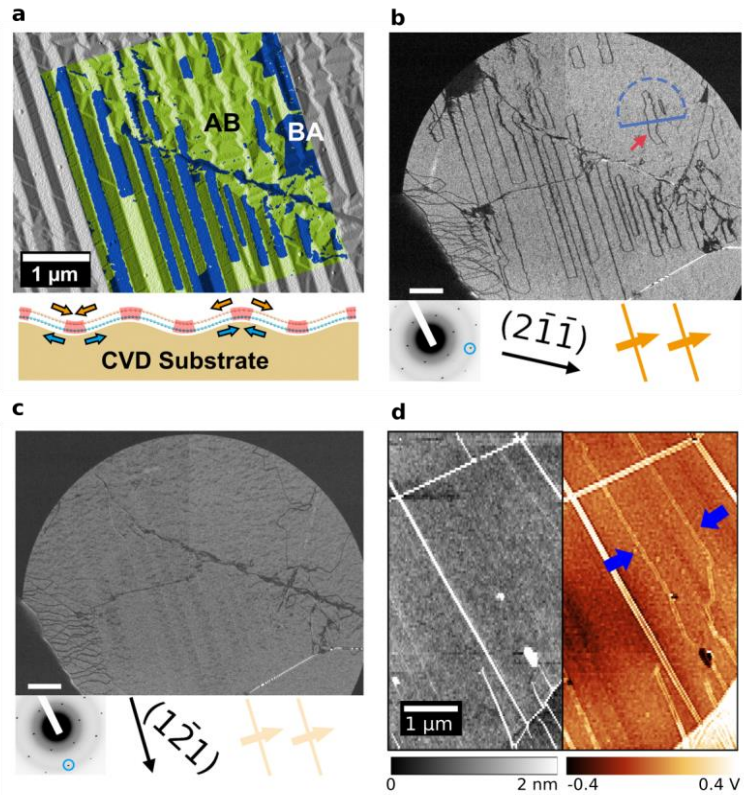

**Figure 1| Substrate corrugation induces proximity-avoiding LSWs. a**, top: Overlay of DF-TEM image from $\{\bar{1}01\}$ reflections with AFM of the underlying catalytic growth substrate. AB (BA) domains identified by DF-TEM are colored blue (green). The grayscale shows the AFM amplitude error signal, where bright (dark) regions correspond to uphill (downhill) corrugation of the substrate. Bottom: Schematic cross-section of CVD bilayer graphene on the growth substrate. Strain intensification sites are colored red, and the orange (blue) arrows represent the strain in the top (bottom) layer of bilayer graphene. **b, c**, top: DF-TEM images of the $\{\bar{2}11\}$ reflections, where the selected diffraction spot is marked by blue circle in the lower parts. The black arrows labeled by Miller indices are vectors corresponding to the selected diffraction spot. The orange lines with arrows indicate the LSWs and the directions of their Burgers vectors, and are invisible (transparent) when perpendicular to the diffraction vector. The scale bar is 500 nm for both (**a** and **c**). In upper part of (**b**), the red arrow indicates a typical LSW loop. Paths going across or around the loop are marked by solid and dashed blue lines, respectively. **d**, AFM topography of another graphene flake is shown on the left and the corresponding near field infrared image is on the right. Parallel LSWs separated by ~1 μm are identified by IR-SNOM, as indicated by the blue arrows.

LSWs formed *via* this mechanism are proximity-avoiding. The coherent nature of the strain across alternating peaks and valleys generates compensating strain in adjacent LSWs. This can be demonstrated by considering an LSW that forms a closed loop (red arrows in Fig. 1b). A path going around the loop experiences zero strain similar to a bilayer with homogeneous stacking order. A path with the same start and end going across the loop must also have zero strain in total, leading to the conclusion that the strain on one side of the loop should compensate the strain in the other. It is worth noting that our situation is totally different from the previous report where the strain is not coherent and atomic defects are required to form a loop[21]. This spatial coherence of the strain associated with this mechanism leads to the appearance of largely parallel LSWs separated by 50 – 200 nm on the growth substrate. The LSW arrangement is, however, metastable. For example, we observed that after transfer onto a carbon film support, LSWs separated by less than 50 nm typically merged/annihilated within a few days (Supplementary Fig. 3e). External mechanical force[14,20] introduced during graphene transfer also led to the annihilation of neighboring LSWs, sometimes leaving well-separated single LSWs as far as 1 µm apart, as imaged by infrared scanning near-field optical microscopy (IR-SNOM; Fig. 1d). The plasmon reflection is large near the LSWs even in regions where the topographic image is featureless[6,13,22]. Proximity-avoiding LSWs were typically found running parallel to wrinkles (bright line features appear in both topography and IR-SNOM images of Fig. 1d), which are also induced by substrate corrugation[23].

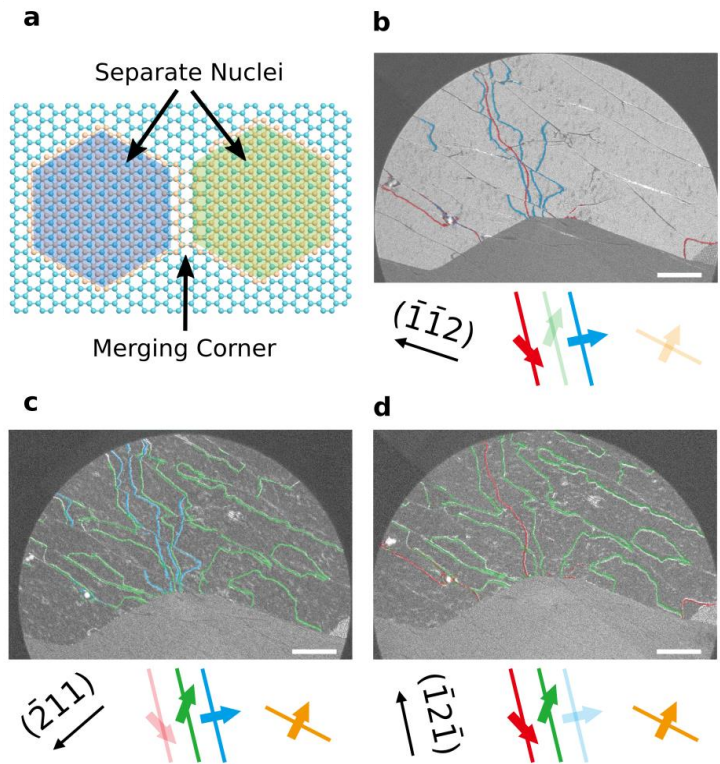

**Figure 2| Close-packed LSWs with random strain. a**, schematic showing the merging corners of two bilayer graphene flakes. The blue and green hexagons represent two second layer regions that nucleated separately during CVD growth. **b-d**, DF-TEM images of a region where two bilayer flakes are merging, selecting $\{\bar{2}11\}$ reflections. A bundle of LSWs runs roughly along the merging interface. Scale bars are 500 nm. Below each image, the black arrows represent the diffraction vectors. The red, green and blue lines with arrows represent the LSWs with different Burgers vectors in the bundle. The yellow line with arrow is for coherent LSWs formed by the substrate corrugation mechanism. Light colored arrows indicate that this Burger's vector direction is not seen in the image due to the invisibility criterion. The LSWs with different Burgers vectors are colored according to the colored arrows.

In contrast to the proximity-avoiding LSWs associated with coherent substrate corrugation, we found that localized stresses act to nucleate close-packed LSW bundles. LSW bundles were associated with contaminant nanoparticles on the substrate (Supplementary Figs. 4a-b), where a large local curvature creates layer asymmetry of in-plane strain in bilayer graphene, as well as with geometrical "point defects" such as the merging corners of two bilayer regions (Fig. 2a). Due to the back-diffusion growth mechanism[18], the top (continuous) layer of the bilayer has a relatively uniform strain distribution, while the merging corners of the two lower layer regions leads to significant stress concentration[24]. This stress concentration and lack of guidance for strain direction leads to dense LSWs with random strain emitting from the stress concentration point. As illustrated in the DF-TEM images (Figs. 2b-d), the Burgers vectors of LSWs induced by substrate corrugation are coherent, as they disappear completely in Fig. 2b, following the invisibility criterion. However, the LSWs nucleated from the merging corners do not disappear at the same time, confirming that they have a distribution of Burgers vectors. We found that these bundles of close-packed LSWs were retained even after transfer onto the silicon substrate (Supplementary Fig. 4c-d), while proximity-avoiding LSWs induced by substrate corrugation were largely annihilated. LSWs that nucleate at stress concentration points can interact with LSWs induced by substrate corrugation, and the separation between incoherent LSWs tends to increase away from the stress concentration points.

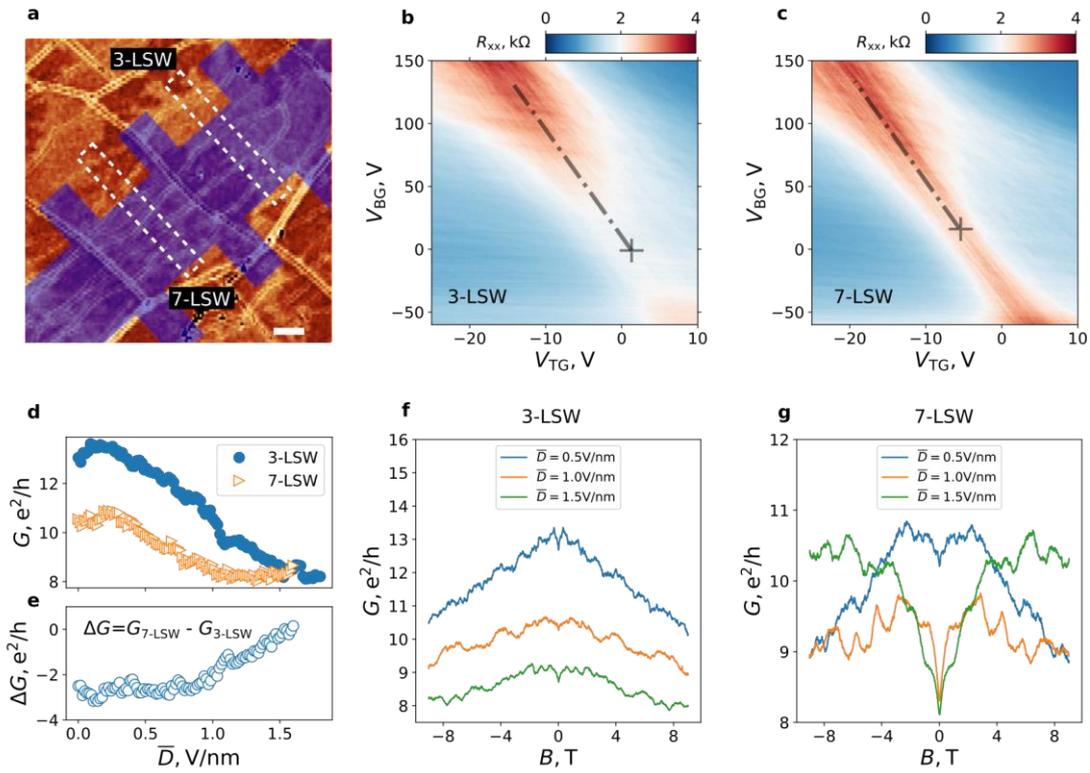

**Figure 3| Electronic Transport of well-separated LSWs and close-packed LSW bundle**. **a**, the IR-SNOM image for neighboring devices of 3 well-separated LSWs and 7 close-packed LSWs. The purple area outlines the device region, and white dashed-line box indicates the top gated area. The scalebar is 500 nm. **b**, **c**, two-dimensional color plot of the longitudinal resistance $R_{xx}$. The dash-dot line denotes the $n = 0$ line cut, and the cross markers indicate the $\bar{D} = 0$ points. **d**, plots of longitudinal conductance against displacement field for dash-dot lines shown in (**b**) and (**c**). **e**, the conductance difference between the two devices, where $\Delta G = G_{7\text{LSW}} - G_{3\text{LSW}}$. **f**, **g**, MC with different displacement field $\bar{D}$ for the 3-LSW and 7-LSW devices, with $n = 0$.

To study the electron transport properties of well-separated and close-packed LSWs, we fabricated pseudo-four-probe devices (Supplementary Note 3). Fig. 3a is a near field infrared image where substrate corrugation-induced, well-separated LSWs interact with stress concentration points, leading to a region with three LSWs separated by ~400 nm within a few micrometers of a close-packed bundle of 7 LSWs. The bilayer graphene is etched to a bar with a width of 1.5 µm and top gates that define a channel length $L = 400\ nm$ for both 3-LSW and 7-LSW devices. The perpendicular electric field and Fermi level can be controlled by the combination of top and bottom gate voltages. The dual gate configuration provides two control knobs, $\bar{D} = (D_b + D_t)/2$ and $n = D_b - D_t$, where $D_b = \varepsilon_b\ (V_b - V_b^0)/d_b$, $D_t = \varepsilon_t\ (V_t^0 - V_t)/d_t$, $\varepsilon_b\ (\varepsilon_t)$ is the dielectric constant of the bottom (top) gate oxide, $d_b\ (d_t)$ the thickness of the bottom (top) dielectric layer, $V_b\ (V_t)$ the bottom (top) gate voltage, and $V_b^0\ (V_t^0)$ the effective offset gate voltage due to doping from the environment. The average displacement field $\bar{D}$ controls the bandgap of bilayer graphene and at low $\bar{D}$ ($\bar{D} < 2$ V/nm), roughly[25] $E_g \sim \bar{D}$, while the Fermi level of the device is controlled by $n$.

Figs. 3b and 3c show the longitudinal resistance versus top and back gate voltages for the 3-LSW and 7-LSW devices in Fig. 3a. The dash-dot lines are along $n = 0$. The displacement field $\bar{D} = 0$ is marked by a cross, and $\bar{D}$ increases going up and to the left in both figures. The resistance at the charge neutrality point increases with $\bar{D}$ for both devices, and for both devices the increase saturates at large $\bar{D}$, similar to what is observed for single LSW devices (Supplementary Note 3), resulting at a value of ~3.2 $k\Omega$ (~8$e^2$/h conductance) for $\bar{D} = 1.6$ V/nm (also in Fig. 3d). In contrast, the resistance of a device without LSW increases well beyond 100 $k\Omega$ without a sign of saturation (Supplementary Fig. 5a). The finite saturated resistance indicates the existence of conducting QVH domain wall channels afforded by LSWs at large bulk bandgap. However, since the 3-LSW device has fewer conducting channels than the 7-LSW device, each channel in the latter contributes ~0.43 conductance of the former. This finding stands in contradiction to earlier reports that multi-LSWs behave as parallel conductors[6], where the conductance per channel should be similar in samples with different numbers of LSWs (also refer to Supplementary Note 3).

By subtracting off the conductance contribution of the bulk, which is in parallel to the LSWs, we see that as the bulk gap increases, the conductance of the 7-LSW device shows a marked increase compared to that of the 3-LSW sample. This is most directly seen by plotting the conductance difference $\Delta G(\bar{D}) = G_{7LSW} - G_{3LSW}$ (Fig. 3e). Two features are identified. First, $\Delta G$ is flat at $\bar{D} < 0.8$ V/nm, where the bulk conductance dominates. This confirms the expectation that the bulk conductance is very similar for the two devices since they are part of the same bilayer flake and located within a few $\mu$m of each other. Second, as $\bar{D}$ increases beyond 0.8 V/nm, $\Delta G$ increases with $\bar{D}$. In this region, the bandgap is large enough to separate the domain wall channels from the bulk electron puddles created by impurities and disorder[26]. Since the contribution of the bulk is similar for the two samples, this increase in $\Delta G(\bar{D})$ implies that the total domain wall channel conductivity for the 7-LSW device is increasing compared to that of the 3-LSW device. This can also be seen directly in the $G(\bar{D})$ data for the devices (Fig. 3d) in the range $1.0$ V/nm $< \bar{D} < 1.6$ V/nm, where $G_{7LSW}$ has a small conductance increase while $G_{3LSW}$ decreases uniformly, reflecting the decrease in bulk conductance. It is worth noting that $\Delta G$ increases by ~3$e^2$/h as $\bar{D}$ goes from 0.8 V/nm to 1.6 V/nm, which we assume reflects the conductance increase of the domain wall channels in the 7-LSW bundle. This change is very significant compared to the ~8 $e^2$/h total conductance at $\bar{D} = 1.6$ V/nm.

MC measurements also reveal a dramatic difference between the well-separated LSWs and the close-packed LSWs. The MC curves of the 3-LSW device taken at different values of $\bar{D}$ (Fig. 3f) show a small positive MC response near $B = 0$, while the MC turns negative at larger $B$. The positive MC at zero field is insensitive to changes in $\bar{D}$. In contrast, for the 7-LSW device (Fig. 3g), the positive

MC grows significantly with $\bar{D}$, developing a deep minimum at $B = 0$. Notably, the MC minimum at $\bar{D} = 1.5\,\text{V}/nm$ has a depth of ~20%. In general, at $n = 0$, the conductance contribution from isolated domain wall modes is insensitive to magnetic field (Supplementary Note 4). For the 3-LSW device, the $\bar{D}$-independent, small positive MC at zero field reflects weak localization in the bulk bilayer graphene, while the negative MC at larger field can be explained by inhomogeneous transport of bilayer graphene in the presence of LSWs (Supplementary Note 5). However, the most salient feature of the data for the 7-LSW device, *i.e.*, the significant deepening of the MC minimum at $B = 0$ with increasing $\bar{D}$, cannot be explained in this framework.

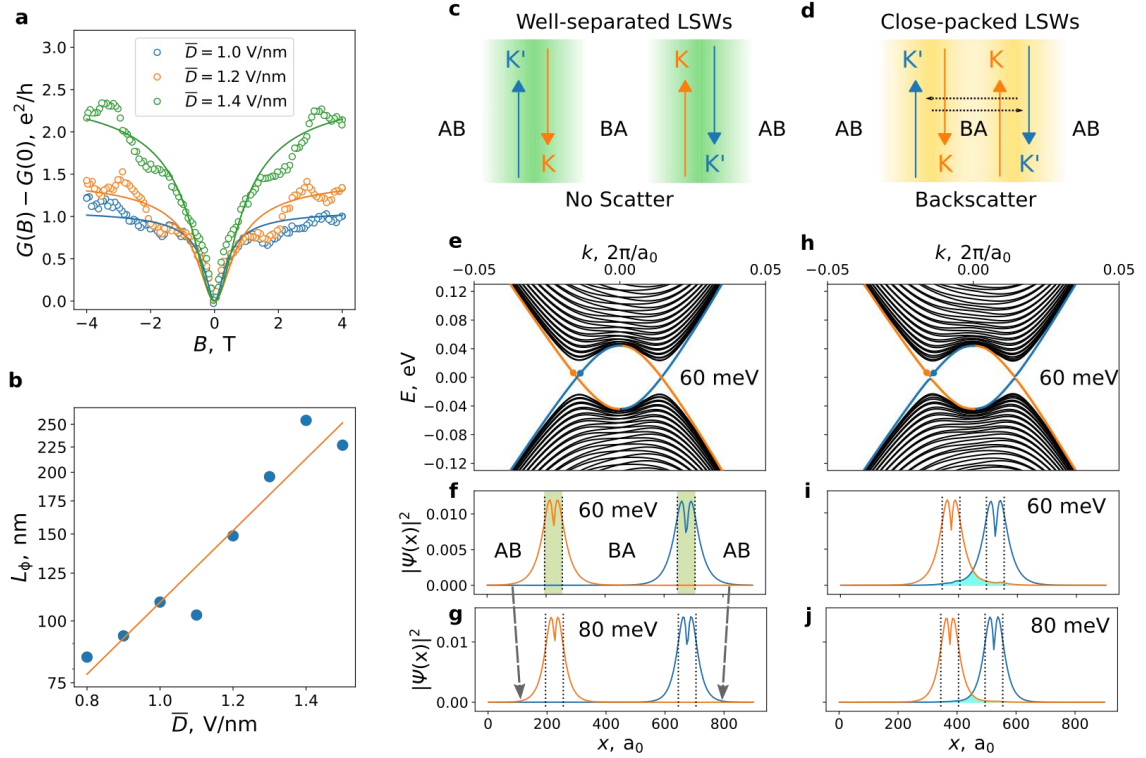

**Figure 4| Theoretical model. a**, three MC curves of the 7-LSW device fitted with 1D weak-localization. The hollow circles mark experimental data points, and the solid lines are the fitted curves. **b**, the coherence length $L_\phi$ variation against $\bar{D}$ in log scale. The orange line is a linear fit of $\ln(L_\phi)$ to the blue dots. **c**, **d**, schematic drawing of the electron transport of well-separated LSWs and close-packed LSWs, respectively. **e**, the band structure calculation of two well separated LSWs in one valley. The bulk bands are black. The orange (blue) bands are mainly localized at the AB/BA (BA/AB) LSW. **f**, the position-space wavefunction envelopes of near mid-gap states localized on different LSWs in the same valley with a 60 meV bulk bandgap. The states in the band structure are indicated by the same color dots in (**e**). The LSW regions are tinted green in-between two dotted lines. The tints are not displayed in (**g, i, j**) for simplicity. **g**, the wavefunction envelopes of near mid-gap states with the same LSW configuration as (**f**) but with an 80 meV bandgap. The gray arrow lines indicate shrinking of the wavefunction envelopes. The distance unit in position space is the monolayer graphene lattice constant $a_0$. **h, i, j,** the band structure of two close-packed LSWs at a 60 meV bandgap (**h**), the corresponding position-space wavefunction envelopes at a 60 meV bandgap (**i**) and an 80 meV bandgap (**j**). The wavefunction overlaps between LSWs are indicated by the light blue patches.

Quantum localization, in general, leads to positive MC responses at zero field. Both 1D and 2D

weak localization theories predict an increased amplitude of positive MC with increasing coherence length $L_\phi$. The MC for 2D weak localization is given by[27]

$$\Delta\sigma(B) = \sigma(B) - \sigma(0) = \frac{e^2}{\pi h}\left[F\left(\frac{B}{B_\phi}\right) - F\left(\frac{B}{B_\phi + 2B_i}\right)\right],$$

where $F(z) = \ln z + \psi\left(\frac{1}{2} + \frac{1}{z}\right)$, $B_{\phi,i} = \frac{\hbar}{4eL_{\phi,i}^2}$, $\psi(z)$ is the digamma function, and $L_i$ is the inter-valley scattering mean free path. The depth of the MC minimum at $B = 0$ increases with $L_\phi$, which is experimentally demonstrated in Ref. [27] and can be understood analytically since $\Delta\sigma(\infty) \sim \ln(1 + 2L_\phi^2/L_i^2)$, where the inter-valley scattering rate is assumed to be constant. On the other hand, the equation for 1D weak localization is[28,29]:

$$G(B) - G(0) = -\frac{2e^2}{h}\frac{N}{L}\left[\left(\frac{1}{L_\phi^2} + \frac{W^2}{3(\hbar/eB)^2}\right)^{-\frac{1}{2}} - L_\phi\right]$$

Where $N$, $W$ and $L$ are the number, width, and length of effective quasi-1D conductors, respectively. The depth of the minimum is $(G(\infty) - G(0)) \sim L_\phi$. Therefore, for the 7-LSW device, the observation of increasing amplitude of the zero-field MC minimum suggests a strong increase in $L_\phi$ with $\bar{D}$, even though as $\bar{D}$ is increased, which increases the energy gap in the bulk, we expect the overall conductance to reflect a transition from a 2D regime dominated by the bulk conductance, to a 1D regime where transport is predominantly via the domain wall channels. For $\bar{D} \geq 0.8$ V/nm, where the 7-LSW domain wall channels show a relative conductivity increase compared with the 3-LSW domain wall channels, we fit the MC using the 1D weak localization equation (Fig. 4a. Additional details are provided in Supplementary Note 6). The extracted $L_\phi$ increases roughly exponentially with $\bar{D}$ (Fig. 4b). Similar results are also found for other close-packed LSW devices (Supplementary Fig. 8).

To understand these measurements, we calculated the electronic Bloch modes for two LSWs (for model details, see Supplementary Note 7). We proposed a model in which domain wall modes remain isolated in well-separated LSWs (Fig 4c) but interact in close-packed LSWs devices where scattering between Bloch states in different LSWs, but with similar crystal momentum can occur (Fig. 4d), influencing the value of both the mean free path and $L_\phi$ of those devices. Moreover, this inter-wall scattering rate varies with the bulk bandgap, making it electrically tunable. Although topological protection suppresses scattering between K and K' valleys, intra-valley scattering between different LSWs is allowed. Because AB/BA and BA/AB LSW are related by spatial inversion, their same-valley modes travel in the *opposite* directions, creating potential backscattering channels whenever these modes overlap.

Walls separated by distances $L \gg v_F\hbar/E_g$ have two topological modes per spin per valley per wall that traverse the energy gap (Fig. 4e). Calculated charge densities of the modes are centered about the LSWs and have exponentially decaying weight away from the walls that changes with the magnitude of the energy gap $E_g$ (Figs. 4f-g). As the wall separation decreases, the wavefunctions of these modes begin to overlap for states slightly above $E = 0$ (Figs. 4h, i), so that transport backscattering is no longer forbidden. This model qualitatively accounts for the observed difference in MC behavior for the two samples: Well-separated domain wall modes contribute additively to the conductivity with a conductance that is insensitive to magnetic fields of this scale, while close packed LSWs possess backscattering channels whose associated transport loops invoke weak localization effects originating from their enclosed magnetic flux.

Magnetotransport responses to electric displacement $\bar{D}$ further support this model. $\bar{D}$ controls localization of domain wall states and inter-wall coupling through $E_g$. The wavefunction of domain wall states localized at a single LSW decays as $e^{-x/x_0}$, where $x$ is the transverse direction to LSW propagation and $x_0 \propto \frac{1}{E_g}$.[9] Since roughly $E_g \sim \bar{D}$, increasing $\bar{D}$ ultimately leads to decreased wavefunction overlap between closely spaced LSWs (Figs. 4i, j), reducing the inter-wall backscattering (Supplementary Fig. 9). If we consider the close-packed LSW device as an effective quasi-1D conductor with multiple coupled domain wall modes, its coherence length and mean free path increase with increasing $\bar{D}$ as a direct result of the decrease of wavefunction overlap. Within this framework, Figure 3e highlights the contribution of interwall coupling on conductivity changes by subtracting the 3-LSW response from the 7-LSW response. The resulting quantity shows the interwall coupling suppresses conductivity at small $\bar{D}$, but that this effect disappears as $\bar{D}$ increases, consistent with a predicted increase in mean free path as $\bar{D}$ is applied. Moreover, the roughly exponential increase of $L_\phi$ with $\bar{D}$ (Fig. 4b) is consistent with an exponential dependence of wavefunction overlap with $\bar{D}$.

In conclusion, we have demonstrated how different strain sources in bilayer graphene can be exploited to create electronically tunable bundles of QVH domain wall modes. Our approach relies upon a scalable, large-area back-diffusion growth process that introduces spatially coherent strain, leading to well-separated LSWs, and localized strain that nucleates close-packed LSW bundles. After transfer onto suitable substrates, the LSWs can be imaged using IR-SNOM so that networks with desired geometry can be electrically contacted for measurement of the transport properties. For the case of a close-packed bundle of LSWs, we demonstrated a change of coherence length as a function of the applied gate voltages and proposed a theoretical model based on the electrically tunable wavefunction overlap of close-packed LSWs. The coexistence of distinct strain sources in this scalable materials system makes it a versatile platform for further demonstrations of programmable quantum transport systems, such as all-electric-controlled valleytronics and Chalker-Coddington type network[30]. Finally, the electrically tunability we have demonstrated for this LSW quantum transport network opens the possibility of QVH state based qubits, similar to how control of interdot coupling enables quantum dot qubits[31].

## Methods

**Bilayer graphene growth and transfer**: The growth method is detailed in ref. 18. Generally, 160 nm Ni was sputtered on one side of Cu foils of 25 µm (Alfa Aesar Item #46365) at 4 mTorr, 100 W (Lesker PVD75 DC/RF Sputter System). Ni-Cu sheets was put into a homemade atmospheric CVD system (the furnace is Lindberg Blue M, Thermo Scientific Co.), ramping to 1050 °C at a rate of 60 °C/min in a flow of 500 sccm Ar and 30 sccm $H_2$. BLG flakes were grown using 2.2 sccm $CH_4$ (1% in Ar) for 1-3 hours. The sample was rapidly cooled down to room temperature with 10 sccm $H_2$ and 1000 sccm Ar after growth.

**DF-TEM:** The graphene flakes were transferred onto an amorphous carbon TEM grid (Electron Microscopy Sciences, CF-400-CU) using a conventional bubbling method (Ref. 18). Centered DF experiments were performed in a JEOL F200 TEM operating at 200 keV and a beam current of 10 nA. Images were acquired with a Gatan OneView camera under a diffraction camera length of 80 cm$^{-1}$ and selected area aperture with effective 1 µm diameter. Tilt angles[32] of up to 20° were used to enhance DF contrast.

**IR-SNOM:** The infrared nano-imaging technique is based on a tapping mode AFM (Bruker Innova). An infrared light at 10.6 µm (MIRcat Daylight Solutions) was focused onto the apex of a metallic AFM tip (Nanoandmore). The enhanced optical field at the tip apex interacts with graphene underneath the tip. A HgCdTe detector placed in the far field is used to collect the scattered light, which carries local optical information of the sample. We demodulated the detected signal at the third harmonic of the tip tapping frequency (Zurich Instrument HF2LI Lock-in Amplifier) to suppress the background contributions. Near-field images are recorded simultaneously with the topography information during the measurements.

**Fabrication of dual-gate device:** After IR-SNOM identification of LSWs, bilayer graphene is patterned by e-beam lithography followed by oxygen plasma etching. The remaining e-beam resist (PMMA 950 A4, MicroChem Corp.) was removed using acetone and isopropanol (IPA). The top contact electrodes for the device (1 nm Cr / 40 nm Pd) were deposited by thermal evaporation following e-beam lithography. After liftoff process, the device was annealed at 225 °C in the forming gases of 1000 sccm Ar and 250 sccm $H_2$ for 1 hour to reduce the PMMA residuals on BLG channels. An HSQ buffer layer (1% in MIBK, 6000 rpm for 60 seconds, softback at 80 °C for 4 minutes) is subsequently spin coated before atomic layer deposition (Cambridge Nanotech S200) of 40 nm $Al_2O_3$ as the top dielectric layer. Lastly, 5 nm Ti/ 40 nm Au top gate was patterned and deposited.

**Electrical transport measurement:** Electrical transport measurements were performed in a Quantum Design Physical Property Measurement System at a temperature of 2K and an ambient He pressure of ~0.1 Torr. Top and bottom gates were independently controlled by Keithley 2410 and Keithley 237 source-measure units, respectively. For each gate, the source-measure unit was placed in a two-probe voltage source configuration with the drain acting as a common ground. The conducting channel was biased and measured using a Stanford Research SR830 lock-in amplifier in a four-probe configuration. With the drain grounded, excitation source was set as 10 mV at 1 kHz, with a 1 MΩ resistor in series to regulate the current to 10 nA. Voltage sense leads were connected to floating A and B inputs to allow lock-in detection of the (A-B) response.

**Data availability:** The data sets generated during the current study, and/or analysed during the current study, are available from the corresponding author upon reasonable request.

# Acknowledgments

This work was supported by the NSF through MRSEC DMR-1720530 and EAGER 1838412. P.M.D., W.M.P., M.D. acknowledge support from NSF through EFRI-1542707 and EAGER-1838456 as well as NIH R21HG010536. J.B.'s contribution was performed under the auspices of the U.S. Department of Energy by Lawrence Livermore National Laboratory under Contract DE-AC52-07NA27344. Z. G. thanks the support from the Research Grant Council of Hong Kong (Project No. 24201020 and 14207421), the National Natural Science Foundation of China (Project No. 62101475), and the Research Matching Grant Scheme of Hong Kong Government (Project No. 8601547). S.W. acknowledges the support by the Fundamental Research Funds for the Central Universities (Grant No. 2042022kf1060). The IR-SNOM measurement was mainly supported by the Director, Office of Science, Office of Basic Energy Sciences, Materials Sciences and Engineering Division of the U.S. Department of Energy under Contract No. DE-AC02-05-CH11231 (sp2-Bonded Materials Program KC2207). Theoretical Modelling by Z.A. and E.J.M. was supported by the Department of Energy Office of Basic Energy Sciences under grant DE FG02-84ER45118. This work was carried out in part at the Singh Center for Nanotechnology, part of the National Nanotechnology Coordinated Infrastructure Program, which is supported by the National Science Foundation grant NNCI-1542153.

## Contributions

## Additional Information

**Supplementary Information** is available for this paper.

Correspondence and requests for materials should be addressed to E.J.M. (mele@physics.upenn.edu), Z.L. (keztluo@ust.hk) and A.T.C.J. (cjohnson@physics.upenn.edu).